\global\def\draftcontrol{0}

   \def\versionno{masterds}

\catcode`\@=11

\expandafter\ifx\csname draftcontrol\endcsname\relax\global\def\draftcontrol{0}
\fi

{\count255=\time\divide\count255 by 60
\xdef\hourmin{\number\count255}
\multiply\count255 by-60\advance\count255 by\time
\xdef\hourmin{\hourmin:\ifnum\count255<10 0\fi\the\count255}}
\def\draftdate{\number\month/\number\day/\number\year\ \ \ \hourmin }

\newcommand\makepapertitle{\par
  \begingroup
    \renewcommand\thefootnote{\@fnsymbol\c@footnote}%
    \def\@makefnmark{\rlap{\@textsuperscript{\normalfont\@thefnmark}}}%
    \long\def\@makefntext##1{\parindent 1em\noindent
            \hb@xt@1.8em{%
                \hss\@textsuperscript{\normalfont\@thefnmark}}##1}%
     \newpage
     \global\@topnum\z@   
     \@makepapertitle
     \thispagestyle{empty}\@thanks
  \endgroup
  \setcounter{footnote}{0}%
  \global\let\thanks\relax
  \global\let\makepapertitle\relax
  \global\let\@makepapertitle\relax
  \global\let\@thanks\@empty
  \global\let\@author\@empty
  \global\let\@date\@empty
  \global\let\@title\@empty
  \global\let\title\relax
  \global\let\author\relax
  \global\let\date\relax
  \global\let\and\relax
  \def\version{\let\version\@version\@gobble}
}
\def\@makepapertitle{%
  \newpage
   \ifnum\draftcontrol=1 {}
   \version\versionno
   \vskip 3em%
   \else
   \hfill\hbox to 3cm {\parbox{4cm}{\@pubnum}\hss}%
   \vskip 3em%
   \fi
   \begin{center}%
   \let \footnote \thanks
     {\LARGE {\@title}}%
     \vskip 1.5em%
     {\normalsize
       \lineskip .5em%
       \begin{tabular}[t]{c}%
         \@author
       \end{tabular}\par}%
     \vskip 1.5em%
     {\@bstract}%
     \end{center}%
     \vskip 1.5em
     \@date%
   \par
}

\gdef\@pubnum{}
\def\pubnum#1{%
  \gdef\@pubnum{#1}}

\gdef\@bstract{}
\def\Abstract#1{%
  \gdef\@bstract{%
   \parbox{\textwidth-0pc}{%
   \centerline{\bf Abstract}\penalty1000%
\kern.2cm%
\noindent
\renewcommand\baselinestretch{1.0}%
{#1}}}
}

\def\ps@paper{\let\@mkboth\@gobbletwo%
     \ifnum\draftcontrol=1
    \def\@oddfoot{\hbox to \textwidth{\tiny \versionno \hfil\tiny\draftdate}%
    \hskip -\textwidth \hbox to \textwidth{\hfil\rm\thepage\hfil}}%
     \else\def\@oddfoot{\hbox to \textwidth{\hfil\rm\thepage\hfil}}
     \fi
     \let\@evenfoot\@oddfoot
}

\def\body{\clearpage
          \pagestyle{paper}
    }

\def\@version#1{\ifnum\draftcontrol=1
\typeout{}\typeout{#1}\typeout{}
\vskip3mm\centerline{\hbox{\fbox{\normalsize{\tt DRAFT -- #1 -- }
                   {\draftdate}}}}\vskip3mm
\fi}
\let\version\@version
\long\def\eqlabel#1{\ifnum\draftcontrol=1
                    \tag@false  
                    \tag*{(\theequation) \hbox to -0.2cm{\hspace{0cm}\small{#1}\hss}}
                    \refstepcounter{equation}
                    \edef\@currentlabel{\theequation}
                    \ltx@label{#1}          
                    \else
                    \label{#1}
                    \fi
                    }
\let\st@bibitem\@bibitem
\let\st@lbibitem\@lbibitem
\ifnum\draftcontrol=1
  \def\@bibitem#1{%
    \st@bibitem{#1}\a@@label{#1}\ignorespaces}
  \def\@lbibitem[#1]#2{%
    \st@lbibitem[#1]{#2}\a@@label{#2}\ignorespaces}
  \def\a@@label#1{%
    \gdef\a@lab{\smash{\normalfont\small#1}}
    \ifvmode
      \if@inlabel
        \global\setbox\@labels\hbox{%
          \llap{\a@lab\let\a@lab\relax
                \kern\@totalleftmargin\kern\marginparsep}%
          \box\@labels}%
      \fi
    \fi}
\fi

\documentclass[12pt,letterpaper]{article}

\usepackage{amsmath,amssymb,array,calc,epsfig,rotating,bm,xcolor}
\usepackage[sort]{cite}
\usepackage{graphicx,esint,float}
\usepackage{psfrag,verbatim}
\usepackage[makeroom]{cancel}
\usepackage{xcolor,url}
\usepackage{hyperref}

\ifnum\draftcontrol=1
\fi

\renewcommand\baselinestretch{1.25}
\renewcommand\section{\@startsection {section}{1}{\z@}%
                                   {-3.5ex \@plus -1ex \@minus -.2ex}%
                                   {2.3ex \@plus.2ex}%
                                   {\normalfont\large\bfseries}}
\renewcommand\subsection{\@startsection{subsection}{2}{\z@}%
                                   {-3.25ex\@plus -1ex \@minus -.2ex}%
                                   {1.5ex \@plus .2ex}%
                                   {\normalfont\normalsize\bfseries}}
\renewcommand\subsubsection{\@startsection{subsubsection}{3}{\z@}%
                                   {-3.25ex\@plus -1ex \@minus -.2ex}%
                                   {1.5ex \@plus .2ex}%
                                   {\normalfont\normalsize\it}}
\renewcommand\paragraph{\@startsection{paragraph}{4}{\z@}%
                                   {-3.25ex\@plus -1ex \@minus -.2ex}%
                                   {1.5ex \@plus .2ex}%
                                   {\normalfont\normalsize\bf}}

\numberwithin{equation}{section}

\def\revise#1       {\raisebox{-0em}{\rule{3pt}{1em}}%
                     \marginpar{\raisebox{.5em}{\vrule width3pt\
                     \vrule width0pt height 0pt depth0.5em
                     \hbox to 0cm{\hspace{0cm}{%
                     \parbox[t]{4em}{\raggedright\footnotesize{#1}}}\hss}}}}

\newcommand\nxt[1]  {\\\fnxt#1}
\newcommand{\ie}{{\it i.e.,}\ }
\newcommand{\eg}{{\it e.g.,}\ }

\def\hp           {{H_{p_j}}}

\def\cala         {{\cal A}}

\def\calm         {{\cal M}}

\def\calo         {{\cal O}}

\def\cals         {{\cal S}}

\def\zet          {{\mathbb Z}}

\def\del          {\partial}

\def\Im   {{\mathfrak{Im}}}

\def\sqr#1#2{{\vcenter{\vbox{\hrule height.#2pt
 \hbox{\vrule width.#2pt height#1pt \kern#1pt
 \vrule width.#2pt}\hrule height.#2pt}}}}

\def\aa1{\phi}
\def\cc1{\psi}

\def\hw{\hat{\omega}}

\def\f0{\text{\boldmath$\varphi$}}
\def\h2{\mathfrak{h}}

\def\qft{{\rm QFT}_{d+1}}

\catcode`\@=12

\begin{document}


\title{\bf Master equations for de Sitter DFPs}

\date{July 19, 2022}

\author{
Alex Buchel\\[0.4cm]
\it $ $Department of Physics and Astronomy\\ 
\it University of Western Ontario\\
\it London, Ontario N6A 5B7, Canada\\
\it $ $Perimeter Institute for Theoretical Physics\\
\it Waterloo, Ontario N2J 2W9, Canada
}

\Abstract{We develop master equations to study perturbative stability
of de Sitter Dynamical Fixed Points (DFPs) of strongly coupled
massive quantum field theories in $d+1$ space-time dimensions with a holographic dual.
The derived spectrum of linearized fluctuations characterizes the late-time
dynamics of holographic strongly coupled non-conformal gauge theories in de Sitter background.
Numerous checks and examples
are presented.
}

\makepapertitle

\body

\version\versionno
\tableofcontents

\section{Introduction and summary}\label{intro}

Dynamical Fixed Points (DFPs) \cite{Buchel:2021ihu} are internal states of interactive quantum
field theories with spatially homogeneous and time-independent one-point
correlation functions of its stress energy tensor $T^{\mu\nu}$ ,
and (possibly additional) set of gauge-invariant local operators $\{\calo_i\}$,
which are characterized by strictly positive divergence of the
entropy current $\cals^\mu$ at late-times $t$,
\begin{equation}
\lim_{t\to \infty} \left(\nabla\cdot \cals\right)\ >\ 0 \,.
\eqlabel{entcurdiv}
\end{equation}
The latter statement implies that a DFP is a genuinely non-equilibrium state of matter.
A prototypical example of a DFP is a late-time state of an interactive non-conformal
field theory in de Sitter space-time\footnote{There is a growing literature on the
subject regarding holographic models \cite{Buchel:2017pto,Buchel:2017qwd,Buchel:2017lhu,Buchel:2019qcq,Buchel:2019pjb,Casalderrey-Solana:2020vls,Ecker:2021cvz,Buchel:2021ihu,Penin:2021sry}.
The understanding of weakly interactive DFPs is still missing --- here, the difficulty is related
to the construction of the far-from-equilibrium entropy current.}. Given
that the Standard Model of particle physics is non-conformal, and we live in
an asymptotically de Sitter space-time \cite{WMAP:2003ivt}, a DFP is the fate
of our Universe.
 
In this paper we develop a compact set of master equations characterizing the spectrum of
linearized fluctuations of holographic  gauge theories in $d+1$ (boundary) de Sitter
space-time dimensions. These equations are akin to the master equations describing the
ring-down, \ie the spectrum of the quasinormal modes (QNMs), of black holes and black branes
\cite{Kodama:2003jz,Jansen:2019wag,Buchel:2021ttt}. While it is physically most intuitive
to describe the fluctuations in the Eddington-Finkelstein (EF) coordinate system of the
gravitational holographic dual, the most compact form of the equations
arises from their implementation in the Fefferman-Graham coordinate system,
with prudent parameterization. As for the black hole QNMs, the correct choice of the
boundary conditions for the DFP fluctuations is vital --- we find a parallel here
with the QNMs boundary conditions in EF and FG coordinate systems.

The spectrum of a black brane QNMs depends on their spatial momentum along the translationary
invariant horizon; in this paper we restrict the discussion
to the DFP fluctuations at zero spatial
momentum\footnote{Finite momentum analysis of the DFP fluctuations in
the model introduced in \cite{Buchel:2017lhu} can be found in
\cite{Penin:2021sry}; see also section \ref{old}.}. This is a relevant
approximation for the late-time dynamics of QFTs in de Sitter, since
any physical momentum is red-shifted to zero, and, a posteriori, we
find that the zero momentum spectrum is gapped\footnote{I would like
to thank David Mateos for a useful discussion regarding this point.}. 

The rest of the paper is organized as follows. We introduce the
holographic model in section \ref{master} and briefly review the set up
of its boundary de Sitter dynamics\footnote{Additional details can be found in
\cite{Buchel:2017pto}.}. We parallel the discussion of a DFP
and its linearized fluctuations in the EF coordinate system (section \ref{efmaster})
and in the FG coordinate system (section \ref{fgmaster}).
Master equations \eqref{fin1}-\eqref{fin3} are presented in a box. 
Section
\ref{examples} contains discussion of a large set of analytic and
numerical examples. The spectra computed in the EF coordinate system
and from the FG coordinate system master equations match perfectly.

Using the derived master equations, it is straightforward to
analyze the stability of the cascading gauge theory de Sitter
DFPs \cite{inprep}.

\section{Master equations}\label{master}

The effective gravitational action dual to a massive $\qft$ is
taken as
\begin{equation}
S_{d+2}=\frac{1}{16\pi G_{d+2}}\int_{\calm_{d+1}}d^{d+2}\xi \sqrt{-g}\biggl[R-\frac 12\sum_{j=1}^p
\left(\del\phi_j\right)^2-V\left(\{\phi_j\}\right)\biggr]\,,
\eqlabel{ea}
\end{equation}
where the bulk scalar fields $\{\phi_j\}$, $j=1\cdots p$, are dual to
the $\qft$ operators $\{\calo_j\}$ of conformal dimensions $\{\Delta_j\}$.
Since our $\qft$ is not a conformal gauge theory, we assume that at least
one of the relevant operators has a nonvanishing source, representing
(at least) one of the mass scales of the boundary $\qft$. 
$V\{\phi_j\}$ is an arbitrary gravitation bulk scalar potential --- its
precise form is not important; if our boundary theory is conformal in
the ultraviolet (UV),
\begin{equation}
V\left(\{\phi_j\}\right)=-\frac{d(d+1)}{L^2}+\frac12 \sum_{j=1}^p\ \underbrace{\frac{\Delta_j(\Delta_j-(d+1))}{L^2}}_{\equiv m_j^2}\ \phi_j^2+\calo(\phi^3)\,,
\eqlabel{uvv}
\end{equation}
where $L$ is the radius of the asymptotic $AdS_{d+2}$ bulk geometry, related to the 
central charge of the UV fixed point. 

We are interested in the dynamics of $\qft$ in de Sitter space-time
\begin{equation}
ds_{d+1}^2=-dt^2+e^{2Ht} d\bm{x}^2\,,
\eqlabel{desitterbackground}
\end{equation}
where $H$ is a Hubble constant.
Following \cite{Buchel:2017pto}, a generic state of the theory,
homogeneous and isotropic in the spatial boundary coordinates
$\bm{x}=\{x_1,\cdots,x_d\}$, leads to a dual gravitational metric ansatz
\begin{equation}
ds_{d+2}^2=2 dt \left(dr-Adt\right)+\Sigma^2\ d\bm{x}^2\,,
\eqlabel{md}
\end{equation}
with the warp factors $A$, $\Sigma$ as well as the bulk scalars
$\phi_j$ depending only on $\{t,r\}$. The equations of motion
obtained from the effective action \eqref{ea} are collected in
appendix \ref{efframe}. They are supplemented with the near-boundary
$r\to \infty$ condition enforcing the background metric
\eqref{desitterbackground}
\begin{equation}
\lim_{r\to \infty}\ \frac{\Sigma^2}{2A}\ =\ e^{2Ht}\,.
\eqlabel{bcmetric}
\end{equation}
An initial state of the $\qft$ is specified providing the
scalar profiles $\phi_j(0,r)$, along with the source terms
(breaking the scale invariance), and solving the constraint
\eqref{con1}, subject to the boundary condition \eqref{bcmetric}.
Equations \eqref{ev1}-\eqref{ev3} are used to holographically evolve the state
for $t>0$. Finally, the constraint \eqref{con2} enforces the
conservation of the boundary $\qft$ stress-energy tensor --- it
requires a single integration constant: the initial energy density of the
state. Implementation details of the outlined de Sitter dynamics
in various models can be found in
\cite{Buchel:2017pto,Buchel:2017lhu,Buchel:2019pjb,Buchel:2021ihu}.

The holographic formulation of the $\qft$ dynamics allows for a natural
definition of its far-from-equilibrium entropy density.
A gravitational geometry \eqref{md} has an apparent horizon located at
$r=r_{AH}$, where \cite{Chesler:2013lia}
\begin{equation}
\left(\del_t+A\del_r\right)\Sigma\bigg|_{r=r_{AH}}=0\,.
\eqlabel{ahloc}
\end{equation}
In a variety of holographic models
\cite{Buchel:2014gta,Buchel:2017pto,Buchel:2019pjb}
it was rigorously proven that the comoving gravitational entropy density of this
apparent horizon, 
\begin{equation}
s^{AH}_{comoving}=\frac{1}{4G_{d+2}}\ \Sigma^d\bigg|_{r=r_{AH}}\,,
\eqlabel{ahent}
\end{equation}
can not decrease with time, \ie
\begin{equation}
\frac{ds^{AH}_{comoving}}{dt}\ \ge\ 0\,,
\eqlabel{dscom}
\end{equation}
leading to its identification with the $\qft$ non-equilibrium comoving entropy
density, 
\begin{equation}
s_{comoving}=\left(e^{Ht}\right)^d\ s(t)\ \equiv\ s^{AH}_{comoving}\,,
\eqlabel{scom}
\end{equation}
where the first equality relates the physical $s$ and the comoving $s_{comoving}$
entropy densities of the $\qft$. Defining the holographic entropy
current\footnote{In a QFT,
the definition of the non-equilibrium entropy current is
ambiguous. This ambiguity is reflected in the dependence of the apparent
horizon on the choice of the spatial slicing in a dual holographic geometry.
At late times, the  slicing must respect the spatial symmetry of a   DFP.
Thus, while the total entropy produced in evolving to a DFP might differ
for different observers, the entropy production rate is
observer independent as $t\to \infty$.} as
\cite{Buchel:2021ihu}
\begin{equation}
\cals^\mu=s(t)\ u^\mu\,,\qquad u^\mu\equiv (1,\underbrace{0,\cdots 0}_d)\,,
\eqlabel{scurrent}
\end{equation}
results in (see \eqref{dscom})
\begin{equation}
\nabla\cdot \cals = \left(e^{-H t }\right)^{d}\ \frac{d}{dt} s_{comoving}\ \ge\ 0 \,.
\eqlabel{ns}
\end{equation}

A DFP is the $t\to \infty$ limit of the dynamical evolution of the system:
\begin{equation}
\lim_{t\to \infty}\ \{A,\phi_j\}(t,r)=\{a,p_j\}(r)\,,\qquad
\lim_{t\to \infty}\ \frac{\Sigma(t,r)}{e^{Ht}}=\sigma(r) \,,
\eqlabel{1pnt}
\end{equation}
and
\begin{equation}
\lim_{t\to\infty} s(t)=s_{ent}\ >\ 0\,.
\eqlabel{slim}
\end{equation}
The existence of the limit \eqref{1pnt} implies that all one-point correlation
functions of the $\qft$ approach a fixed value at late times, while the limit
\eqref{slim} implies that
\begin{equation}
\lim_{t\to\infty}\ \left(\nabla\cdot \cals\right)= d\ H\ s_{ent}\ >\ 0\,,
\eqlabel{entprod}
\end{equation}
\ie at  late-times the comoving entropy production rate approaches a
constant.

\subsection{The Eddington–Finkelstein coordinates}\label{efmaster}

It is convenient to introduce dimensionless radial coordinate $x$ and time $u$ as
\begin{equation}
x\equiv \frac{H L^2}{r}\,,\qquad u\equiv H t\,,
\eqlabel{xuef}
\end{equation}
and redefine  (see Appendix \ref{efframe} for the EF frame metric ansatz)
\begin{equation}
A(t,r)\equiv L^2H^2\ A(u,x)\,,\qquad \Sigma(t,r)\equiv L\ e^u\ \sigma(u,x)\,,\qquad
\phi_j(t,r)\equiv \phi_j(u,x)\,,
\eqlabel{rescale}
\end{equation}
to scale-out $L$ and $H$ dependence from \eqref{ev1}-\eqref{con2}.

The most intuitive way to study the spectrum of linearized fluctuations about the DFP
\eqref{1pnt} is to use the EF coordinate system \eqref{md}.
Following \cite{Buchel:2017lhu}, we set
\begin{equation}
\begin{split}
&A(u,x)=\frac{1}{2x^2}\biggl(g(x)+H_1(x)\cdot e^{-i\hw u}\biggr)\,,\qquad \sigma(u,x)=\frac 1x \biggl(f(x)+ H_2(x)\cdot e^{-i\hw u}\biggr)\,,\\
&\phi_j(u,x)=p_j(x)+H_{p_j}(x)\cdot e^{-i\hw u}\,,
\end{split}
\eqlabel{defflef}
\end{equation}
where $\hw$ is the spectral frequency of the linearized fluctuations,
$\{H_1,H_2,H_{p_j}\}$
\begin{equation}
\hw\ \equiv\ \frac wH\ \equiv -i\gamma\,,\qquad {\rm correspondingly,}\qquad \gamma\equiv i\hw \,.
\eqlabel{defwh}
\end{equation}

Substituting \eqref{defflef} into \eqref{ev1}-\eqref{con2} and linearizing in $\{H_1,H_2,\hp\}$, we find
(now $'\equiv \frac{d}{dx}$)
\nxt the background equations:
\begin{equation}
\begin{split}
0=&p_j''-\frac{- x g f'd + f (-g' x +  (x + g)d)}{x f g}\ p_j'-\frac{1}{g x^2}\ \del_j V\,,
\end{split}
\eqlabel{efbac1}
\end{equation}
\begin{equation}
\begin{split}
0=&f''+\frac {f}{2d}\  \sum_{j=1}^p (p_j')^2\,,
\end{split}
\eqlabel{efbac2}
\end{equation}
\begin{equation}
\begin{split}
0=&-2d (d - 1)  x^2 g  (f')^2 + 2 d\left(- g' x +  2(x + g)d\right) f  x  f'+f^2 x^2 g\
\sum_{j=1}^p (p_j')^2\\
&+2 f^2 ( x g'd -d (d + 1) g - 2 xd^2 -V)\,,
\end{split}
\eqlabel{efbac3}
\end{equation}
\begin{equation}
\begin{split}
0=&f^2 x^2 g\ \sum_{j=1}^p (p_j')^2 -2 d (d - 1) x^2 g (f')^2+2 (- g' x +  2gd + 2 x(d - 1))
x f f' d-2 f^2 V\\
&-\frac{2 d f^2}{g} (- x (x + g) g' + (d + 1) g^2 + 2g  xd - 2x^2)\,,
\end{split}
\eqlabel{efbac4}
\end{equation}
\begin{equation}
\begin{split}
0=&g''-\frac{g d (d - 1)}{f^2}\ (f')^2+\frac{2 d (d-1)(x + g)}{x f}\ f'+\frac g2\
\sum_{j=1}^p (p_j')^2-\frac{d - 2}{x^2 d}\ V\\
&+\frac{1}{x^2}(-2 g' x + (-d^2 + d + 2) g - 2 d  (d - 1)x)\,;
\end{split}
\eqlabel{efbac5}
\end{equation}
\nxt the equations for the fluctuations:
\begin{equation}
\begin{split}
&0=\hp''+\frac{x (d + 2) f' - f \left(d
+ \frac{x (d - 2 \gamma + 2)}{\colorbox{yellow}{g}}\right)}{f x}\ \hp'
+\frac{f' \gamma x^2 d - f (x \gamma d+\del^2_{jj}V)}{f x^2 \colorbox{yellow}{g}}\ \hp
\\&-\frac{1}{\colorbox{yellow}{g} x^2}\ \sum_{k\ne j}  H_{p_k} \del^2_{jk}V
+p_j' \left(\frac{H_1'}{\colorbox{yellow}{g}}+\frac{H_2'd}{f}\right)
+\frac{(-2f' g + (d + 2) f) x^2 p_j' + f p_j \del^2_{jj}V}{fx^2 \colorbox{yellow}{g}^2}\ H_1
\\&+ p_j' d\left(-\frac{f'}{f^2} + \frac{\gamma}{f \colorbox{yellow}{g}}\right)\ H_2\,,
\end{split}
\eqlabel{effl5}
\end{equation}
\begin{equation}
\begin{split}
&0=g\ \sum_{j=1}^pp_j' \hp'-\frac{1}{x^2}\ \sum_{j=1}^p p_j \hp \del^2_{jj}V
+\left(\frac{H_1}{2} + \frac{g H_2}{2f}\right)\ \sum_{j=1}^p(p_j')^2-\frac{(f' x - f)d}{f x}\  H_1'\\
&+\frac{2 (- x g f'd + ( g d + x (d - \gamma + 1)) f) d}{f^2 x}
\ H_2'
-\frac{VH_2}{f x^2}+\frac{d (d-1)(-f H_1 + g H_2)}{f^3}\ (f')^2\\&
-\frac{2 d (- H_1 f d+ H_2 (d x \gamma - x \gamma - g))}{x f^2}\ f'
+\frac{ d}{fx^2} \biggl(
-H_1 (d + 1) f + H_2 ( -( d + 1) g \\&+ 2(-1 + (\gamma - 1) d) x)
\biggr)\,,
\end{split}
\eqlabel{effl2}
\end{equation}
\begin{equation}
\begin{split}
&0=g \sum_{j=1}^p \biggl\{
p_j' \hp' +\hp \left(p_j' \gamma-\frac{p_j}{x^2} \del^2_{jj}V
\right)
\biggr\}+\left(H_1+\frac{g H_2}{2f}\right) \sum_{j=1}^p (p_j')^2
-\frac{V}{f x^2} \biggl(H_2\\
&+\frac{f H_1}{g}\biggr)
-\frac{d (g f' x-g f-f x)}{f^2 x g} \left(H_1'f +2d gH_2'\right)
-\frac{d (2 d H_1 f-H_2 g d+g H_2)}{f^3}\ (f')^2
\\&+\frac{2df'}{f^2x} \biggl(H_2 g + (2 d + 1) H_1 f + H_2 x (1-d \gamma ))
+  \frac{(2d + \gamma)xH_1 f}{2g}\biggr)
- \biggl(2(d + 1) \\
&+ \frac{x (2d +  \gamma + 2)}{g}\biggr)\frac{H_1d}{x^2}
+\frac{H_2d}{fx^2} \biggl(-(d + 1) g + 2(-1 + (\gamma - 1) d) x
+ \frac{2\gamma x^2 (\gamma - 1)}{g}\biggr)\,,
\end{split}
\eqlabel{effl3}
\end{equation}
\begin{equation}
\begin{split}
&0=H_1''-\frac2x\ H_1'+g\sum_{j=1}^p p_j' \hp'
+\frac{2 (d - 1) (-g f' x + f (x + g)) d}{x f^2}\ H_2'
+\frac{H_1}{2}\sum_{j=1}^p (p_j')^2\\&
+\sum_{j=1}^p\hp \left(p_j' \gamma-\frac{(d - 2) p_j}{x^2d} \del^2_{jj}V\right)
-\frac{d (d - 1)(f H_1 - 2 g H_2) }{f^3} (f')^2
-\frac{2f' (d - 1) d}{f^2x} (-f H_1 \\
&+ (g + (\gamma + 1) x) H_2)
+\frac{1}{fx^2}(-H_1 (d + 1) (d - 2) f + 2 d \gamma x H_2 (d - 1))\,,
\end{split}
\eqlabel{effl4}
\end{equation}
\begin{equation}
\begin{split}
0=&H_2''+\frac fd\ \sum_{j=1}^p \hp' p_j'+\frac{H_2}{2d}\ \sum_{j=1}^p(p_j')^2 \,.
\end{split}
\eqlabel{effl1}
\end{equation}

Some comments are in order:
\begin{itemize}
\item The radial coordinate $x$ ranges as
\begin{equation}
x\in (0,x_{AH}]\,,
\eqlabel{xrange}
\end{equation}
where $x\to 0$ is the location of the asymptotic AdS$ _{d+2}$ boundary, and
$x_{AH}$ is the location of the apparent horizon, corresponding to $r_{AH}$ in
\eqref{ahloc}. Note that
\begin{equation}
\begin{split}
0=\left(\del_t + A \del_r \right) \Sigma\bigg|_{r=r_{AH}} &= e^u\  L\ \biggl(H+ L^2H^2
\underbrace{\left(-\frac{x^2}{HL^2}\right)}_{\frac {dx}{dr}}A(u,x)\ \del_x\biggr)\ \sigma(u,x)
\bigg|_{x=x_{AH}}  
\\
&=e^u\ LH\ \biggl(1+A(u,x)\ (-x^2 \del_x)\biggr)\ \sigma(u,x)\bigg|_{x=x_{AH}}  \\
&\propto e^u\ \frac{(2x+g)f-g f' x}{2x^2} \bigg|_{x=x_{AH}}\qquad {\rm as}\qquad u\to \infty\,,
\end{split}
\eqlabel{defxah}
\end{equation}
where we used \eqref{xuef} and \eqref{rescale}. Following
\cite{Buchel:2017pto}\footnote{See Appendix A there --- the argument has a straightforward
generalization to the model \eqref{ea}.},
\begin{equation}
(-\del_x)\ \sigma(u,x)>0 \,,
\eqlabel{dersigma}
\end{equation}
thus $x_{AH}$ {\it must} occur at value of $x_{AH}>x^\star$, such that
\begin{equation}
A(u,x)\bigg|_{x=x^\star}=0\,.
\eqlabel{xstar}
\end{equation}
To put it differently, while for equilibrium black brane horizons the radial coordinate ranges
from the asymptotic infinity to the vanishing of $g_{tt}$ components of the metric, \ie the
analogue of $x^\star$, for the holographic DFP realization the radial coordinate
must be extended beyond $x^\star$, heuristically, {\it inside the black brane horizon}.
\item The adopted Eddington–Finkelstein frame has  residual diffeomorphisms
\begin{equation}
r\ \to\ \hat{r}\equiv r+\lambda(t)\qquad \Longleftrightarrow\qquad \frac{1}{x} \to\ \frac{1}{\hat{x}}
\equiv \frac 1x+\lambda(u)\,,
\eqlabel{resrx}
\end{equation}
where $\lambda$ is an arbitrary function of $t$ (or $u$). The transformation \eqref{resrx}
arbitrarily shifts the location of $r_{AH}$ ( $x_{AH}$ ), and must be fixed --- irrespectively
the causal domain for the evolution requires the extension of the radial coordinate
as in \eqref{xrange}. For numerical simulations, it is convenient \cite{Chesler:2013lia}
to fix \eqref{resrx} so that $x_{AH}=1$. To study the DFP perturbations it is instead more
convenient \cite{Buchel:2017lhu} to fix  \eqref{resrx} requiring that the location of
$x^*$ is kept fixed. Thus, here we adopt
\begin{equation}
x^\star=\frac 13\,.
\eqlabel{xsc}
\end{equation}
The choice \eqref{xsc} is motivated by the pure-AdS$ _{d+2}$ solution of the system
\eqref{efbac1}-\eqref{efbac5}:
\begin{equation}
p_j\equiv 0\,,\qquad f=1-x\,,\qquad g=(1-3 x)(1-x) \,,
\eqlabel{pureads}
\end{equation}
which from \eqref{defxah} implies that\footnote{Observe that
$x_{AH}>x^\star$.} $x_{AH}=1$.

It is important to keep in mind that fixing the residual diffeomorphism as in
\eqref{xsc} implies a constraint both for the DFP background function $g(x)$ and the
linearized fluctuation $H_1$. Specifically, from \eqref{defflef},
\begin{equation}
A(u,x)\bigg|_{x=x^\star}=0\qquad \Longrightarrow\qquad g(x^\star)=0\ \&\ H_1(x^\star)=0\,.
\eqlabel{constxs}
\end{equation}
\item The set of equations \eqref{efbac1}-\eqref{efbac5} is redundant: using \eqref{efbac3}
and \eqref{efbac4} we can algebraically eliminate $g'$ and $g$,
\begin{equation}
\begin{split}
g'=&\frac{1}{2 d x f (x f'-f)} \biggl(
g f^2 x^2\ \sum_{j=1}^p (p_j')^2-2 f^2 V
-2 d (d-1)x^2 g  (f')^2\\
&+4 f d^2 x (x+g) (f')-2 f^2 d (g d+2 x d+g)\biggr)\,,\\
g=&\frac{2 f}{f^2 x^2 \sum_{j=1}^p (p_j')^2-2 d (d+1)(x f'-f)^2 }
\biggl(
f V-2 x d (d+1) (x f'-f)
\biggr)\,.
\end{split}
\eqlabel{gpg}
\end{equation}
Thus, as a complete set, one can use \eqref{efbac1} and \eqref{efbac2}, where $g'$ and $g$
are eliminated\footnote{It can be explicitly verified that all the redundant equations
are consistent.} as in \eqref{gpg}, precisely as it was used in \cite{Buchel:2017lhu}.
\item There is an additional first-order constraint, consistent with \eqref{gpg},
\begin{equation}
g'=\frac{2}{f}\ \left(f' g - f\right)\,.
\eqlabel{gpn}
\end{equation}
It is dictated by the consistency of the DFP description in the Eddington-Finkelstein
and the Fefferman-Graham coordinate systems \cite{Buchel:2017pto}.
Indeed, as $t\to \infty$, the EF and the FG DFP holographic geometries take form (compare
\eqref{ef1} and \eqref{fg1})
\begin{equation}
\begin{split}
&{\rm EF:}\qquad ds_{d+2}^2=2 dt\biggl(dr-A(r)\ dt\biggr)+\sigma(r)^2 e^{2 H t}\ d{\bm{x}}^2\,,\\
&{\rm FG:}\qquad ds_{d+2}^2=c_1(\rho)^2\ \biggl(-d\tau^2+e^{2H\tau}\ d{\bm{x}}^2 \biggr)
+c_3(\rho)^2\ d\rho^2\,.
\end{split}
\eqlabel{effgframe}
\end{equation}
The EF time $t$ and FG time $\tau$ are related as
\begin{equation}
dt=d\tau+\frac{1}{2A(r)}\ dr\qquad \Longrightarrow\qquad t=\tau-\int_r^\infty \frac{dz}{2A(z)} \,,
\eqlabel{ttau}
\end{equation}
leading to
\begin{equation}
{\rm EF}\to {\rm FG}:\qquad ds_{d+2}^2= -\underbrace{2A(r)}_{c_1(\rho)^2}\ d\tau^2+
\underbrace{\sigma(r)^2
e^{-2H \int_r^\infty \frac{dz}{2A(z)} }}_{c_1(\rho)^2}\ e^{2H\tau}\ d{\bm x}^2+\frac{1}{2A(r)}\ dr^2\,,
\eqlabel{ef2fg}
\end{equation}
where we highlighted the $g_{\tau\tau}$ and $g_{{\bm x}{\bm x}}$ components of the metric.
Thus, the following identify must be true:
\begin{equation}
\begin{split}
&2A(r)\ \equiv\ \sigma(r)^2 e^{-2H\int_r^\infty\frac{dz}{2 A(z)}}\qquad
\Longleftrightarrow\qquad \frac{d}{dr} \ln\biggl[\frac{2 A(r)}{\sigma(r)^2}\biggr]
-\frac{H}{A(r)}\equiv 0\,,\\
&\Longrightarrow\qquad \frac{x^2}{H L^2 fg}\biggl(2(f'g -f) -fg'\biggr)\equiv 0\,,
\end{split}
\eqlabel{identity}
\end{equation}
where in the second line we used \eqref{xuef} and \eqref{rescale}.
The last identity in \eqref{identity} is precisely \eqref{gpn}.
\item The set of equations \eqref{effl5}-\eqref{effl1} is redundant:
we can use \eqref{effl5}
(as the set of the second-order equations for $\hp$),
as well as \eqref{effl2} and \eqref{effl3} as a pair of the first-order equations for
$H_1$ and $H_2$. We explicitly verified that the redundant equations \eqref{effl4}
and \eqref{effl1} are consistent. Again, this choice of the minimal set of the
fluctuation equations was used in \cite{Buchel:2017lhu}.
\item Notice that the equations for the fluctuations
are singular at $x=x^\star$ --- in \eqref{effl5} we highlighted the dependence on
$\colorbox{yellow} g$, which vanishes precisely at $x=x^\star$, see \eqref{constxs}.
To solve for the spectrum of the linearized fluctuations about the
DFP given by \eqref{efbac1} and \eqref{efbac2} (together with \eqref{gpg})
we impose \cite{Buchel:2017lhu}:
\nxt the normalizability conditions on $\{\hp,H_1,H_2\}$
at the asymptotic AdS$ _{d+2}$ boundary;
\nxt the non-singularity of solution $\{\hp,H_1,H_2\}$ at $x=x^\star$;
\nxt the residual diffeomorphism condition $H_1(x^\star)=0$, see \eqref{constxs}.\\
The above 3 conditions are enough to unambiguously fix the spectrum $\{\gamma\}$
of the linearized fluctuations about the DFP\footnote{The reader
is encouraged to follow the detailed realization of the above computational
framework in the specific model of \cite{Buchel:2017lhu}.}.
\item We would like to stress an important distinction between the computation of
the spectrum of a black brane quasinormal modes and a DFP  linearized fluctuations:
\nxt in the former case, one requires the normalizability of the quasinormal
modes at the asymptotic AdS$ _{d+2}$ boundary, and the regularity of these
modes at the black brane horizon (in the EF coordinate system); 
\nxt in the DFP fluctuation case,  one also requires the
 normalizability of the linearized fluctuations  at the asymptotic AdS$ _{d+2}$ boundary;
 however, the bulk regularity condition is imposed at the location where the $g_{tt}$
 component of the bulk metric vanishes, which is {\it outside} the location of the apparent
 horizon, \ie at $x=x^\star< x_{AH}$.
In simple holographic models, \eg \cite{Buchel:2017lhu} and
\cite{Buchel:2021ihu}, it was explicitly
shown (via numerical simulations)
that the dynamical evolution towards a DFP in EF coordinates is
well approximated by the $t\to \infty$ DFP fluctuations.
The EF frame characteristic-formulation  (equivalently
the evolution that guarantees the incoming-wave boundary condition
on the AH) is governed by equations
\eqref{ef1}-\eqref{con2}, and evidently is always
smooth at any finite time. The $x=x^*$
coordinate singularity arises only when one linearizes these evolution
equations about the strict $t\to \infty$ DFP geometry, as determined
by \eqref{1pnt}. Thus, it is natural to
require\footnote{In \cite{Penin:2021sry} it was
shown that  DFP fluctuations computed
in this manner appear as poles in
the two-point correlation functions
of the boundary stress-energy tensor.} the non-singularity of
fluctuations over all the causal domain --- from the asymptotically
AdS$ _{d+2}$ boundary to the AH. Regularity of the linearized fluctuations
 at $x=x^\star$ guarantees the regularity at $x=x_{AH}$, see \cite{Buchel:2017lhu}.\\
 Above observation implies that there must be a self-consistent procedure
 to extract the spectrum of a DFP linearized fluctuations from the
 part of the geometry $x\in (0,x^\star]$, which has a simple FG coordinate
 system representation. Understanding the computation of a DFP fluctuation
 spectrum in the FG coordinate system will lead us to the set of master equations,
 presented in section \ref{fgmaster}.
\end{itemize}

\subsection{The Fefferman-Graham coordinates}\label{fgmaster}
It is convenient to introduce dimensionless radial coordinate $x$ and time $u$ as
\begin{equation}
x\equiv {H L^2}\ {\rho}\,,\qquad u\equiv H \tau\,,
\eqlabel{xufg}
\end{equation}
and redefine (see Appendix \ref{fgframe} for the FG frame metric ansatz)
\begin{equation}
\begin{split}
&h(\tau,\rho)\equiv L^{4}\ h(u,x)\,,\qquad G_{tt}(\tau,\rho)\equiv G_{tt}(u,x)\,,\qquad
\qquad G_{xx}(\tau,\rho)\equiv G_{xx}(u,x)\,,\\
&\phi_j(\tau,\rho)\equiv \phi_j(u,x)\,,
\end{split}
\eqlabel{rescalefg}
\end{equation}
to scale-out $L$ and $H$ dependence from \eqref{eq1}-\eqref{eqc2}.
To study the spectrum of linearized fluctuations about the DFP
we set
\begin{equation}
\begin{split}
&h(u,x)=h(x)+H_h(x)\cdot e^{-i\hw u}\,,\qquad \sqrt{G_{tt}(u,x)}=1+ H_1(x)\cdot e^{-i\hw u}\,,\\
&\sqrt{G_{xx}(u,x)}=e^u\biggl(1+ H_2(x)\cdot e^{-i\hw u}\biggr)\,,\qquad \phi_j(u,x)=p_j(x)+H_{p_j}(x)\cdot e^{-i\hw u}\,,
\end{split}
\eqlabel{defflfg}
\end{equation}
where $\hw$ is the spectral frequency of the linearized fluctuations as in \eqref{defwh}.

Substituting \eqref{defflfg} into \eqref{eq1}-\eqref{eqc2} and linearizing in
$\{H_h,H_1,H_2,\hp\}$, we find
\nxt the background equations:
\begin{equation}
\begin{split}
0=&p_j''-\frac{x (d + 2) h' + 4 h d)}{4 x h}\ p_j'-\frac{h^{1/2}}{x^2}\ \del_jV\,,
\end{split}
\eqlabel{fgbac1}
\end{equation}
\begin{equation}
\begin{split}
0=&\left(8x^2 \sum_{j=1}^p (p_j')^2  - 16 d (d+1)\right) h^{5/2} - 8 d (d+1 )h'x  h^{3/2}
+ 16 d (d+1)x^2 h^{7/2} \\&- d(d+1) (h')^2 x^2  h^{1/2}-16 h^3 V\,,
\end{split}
\eqlabel{fgbac2}
\end{equation}
\begin{equation}
\begin{split}
0=&h''-\frac{d + 11}{8h}\ (h')^2-\frac{d + 1}{x}\ h'-\frac hd\ \sum_{j=1}^p (p_j')^2
-\frac{2 h^{3/2}}{ x^2d}\ V+\frac{2 h (x^2 (d - 1) h - d - 1)}{x^2}\,;
\end{split}
\eqlabel{fgbac3}
\end{equation}
\nxt the equations for the fluctuations:
\begin{equation}
\begin{split}
&0=\hp''+\frac{-x (d + 2) h' - 4 h d}{4x h}\ \hp'+p_j' (H_1'+H_2' d)
-\frac{p_j' (d + 2)}{4h}\ H_h'\\
&+H_h \biggl(
\frac{p_j'h' (d + 2)}{4h^2} -\frac{1}{2 x^2 h^{1/2}}\ 
\del_jV\biggr)
+\hp \biggl(-\frac{h^{1/2}}{x^2}\ \del^2_{jj}V+h \gamma (d - \gamma)
\biggr)\\
&-\frac{h^{1/2}}{x^2}\ \sum_{k\ne j} H_{p_k} \del^2_{jk}V \,,
\end{split}
\eqlabel{fgfl5}
\end{equation}
\begin{equation}
\begin{split}
&0=-4 h \gamma xd\ H_h'+16 h^2  xd\ H_1'+16 h^2  x d(\gamma - 1)\ H_2'+\gamma d
(5 x h' + 4 h)\ H_h\\
&+8 h^2 \gamma x\ \sum_{j=1}^p \hp p_j' \,,
\end{split}
\eqlabel{fgfl2}
\end{equation}
\begin{equation}
\begin{split}
&0=4 x h d (d + 1) (x h' + 4 h)\ H_h'-16 h^2 d x (x h' + 4 h)\ (H_1'+ H_2'd)
-32 x^2 h^3\ \sum_{j=1}^p p_j'  \hp'
\\&+H_h \biggl(
-40 x^2 h^2\ \sum_{j=1}^p (p_j')^2+96 h^{5/2} V
-16 \biggl(-\frac{x^2 (d + 1)}{16}\ (h')^2 - \frac{3x h (d + 1)}{2}\ h'
\\
&+ h^2 (((\gamma + 7) d - \gamma^2 + 7) x^2 h - 5 d - 5)\biggr) d\biggr)
+H_1 \biggl(
-48 h^3 x^2\ \sum_{j=1}^p (p_j')^2+96 h^{7/2} V
\\&-32 h d \biggl(-\frac{3x^2 (d + 1)}{16}\ (h')^2 - \frac{3x h (d + 1)}{2}\ h'
+ h^2 (x^2 (d + 1 + 2 \gamma) h - 3 d - 3)\biggr)\biggr)
\\&+32 h^{7/2}\ \sum_{j=1}^p \del_jV \hp
+H_2 \biggl(
-32 h^3 x^2\ \sum_{j=1}^p (p_j')^2+64 h^{7/2} V+64 \biggl(
\frac{x^2 (d + 1)}{16}\ (h')^2
\\&+ \frac{x h (d + 1)}{2}\ h'
+ h^2 (x^2 ((\gamma - 1) d - \gamma^2 + \gamma - 1) h + d + 1)\biggr) h d
\biggr)\,,
\end{split}
\eqlabel{fgfl3}
\end{equation}
\begin{equation}
\begin{split}
&0=H_h''-\frac{4 h}{d}\ H_1''-\frac{4 h (d - 1)}{d}\ H_2''
+\frac{x (d + 2) h' + 4 h d}{d x}\ ((d-1)H_2'+H_1')
\\&
+ \frac{(-x (d + 11) h' - 4 h (d + 1))}{4x h}\ H_h'-\frac{2 h}{d}\ \sum_{j=1}^p
p_j' \hp'
+H_h \biggl(
-\frac 1d\ \sum_{j=1}^p (p_j')^2-\frac{3 h^{1/2}}{ x^2d}\ V
\\&+\frac{1}{8 x^2 h^2d} \biggl(
d x^2 (d + 11)\ (h')^2
+ 8 h^2 (x^2 ((\gamma + 4) d^2 + (-\gamma^2 - 3 \gamma - 4) d
+ 2 \gamma^2 \\&+ 2 \gamma) h - 2 d^2 - 2 d)\biggr)
\biggr)
-\frac{2 h^{3/2}}{d x^2}\ \sum_{j=1}^p \del_jV \hp-\frac{4 h^2 (d - 1)
(d - \gamma)}{d} (H_2 \gamma + H_1)\,,
\end{split}
\eqlabel{fgfl4}
\end{equation}
\begin{equation}
\begin{split}
&0=H_h''-4 h\ H_2''+ \frac{-x (d + 11) h' - 4 h (d + 1)}{4x h}\ H_h'
+\frac{x (d + 2) h' + 4 h d}{x}\ H_2'\\&-\frac{2 h}{d}\ \sum_{j=1}^p p_j' \hp'
+H_h \biggl(
-\frac1d\ \sum_{j=1}^p (p_j')^2-\frac{3 h^{1/2}}{ x^2d}\ V
+\frac{1}{8h^2 x^2} \biggl(x^2 (d + 11) (h')^2
\\&+ 8 h^2 (x^2 ((\gamma + 4) d - 2 \gamma - 4) h - 2 d - 2)\biggr)
\biggr)
-\frac{2 h^{3/2}}{ x^2d}\ \sum_{j=1}^p \hp \del_jV\\
&-4 h^2 (H_2 \gamma + H_1) (d - 1)\,.
\end{split}
\eqlabel{fgfl1}
\end{equation}
A careful examination of \eqref{fgfl5}-\eqref{fgfl1} reveals that $\{H_1,H_2\}$ can be eliminated algebraically as
\begin{equation}
H_1=H_2=\frac{(d-2) H_h}{4(d-1) h}\,.
\eqlabel{mgauge}
\end{equation}

\bigskip

\noindent\fbox{%
    \parbox{\textwidth}{%
{
The remaining equations can be recast in {\it masters'} form:
\begin{equation}
\begin{split}
&0=\hp'' + \frac{-x (d + 2) h' - 4 h d}{4xh}\ \hp'
+\hp \biggl(-\frac{h^{1/2}}{x^2}\ \del^2_{jj}V + h \gamma (d - \gamma)\biggr)
\\&- \frac{h^{1/2}}{x^2}\ \sum_{k\ne j} H_{p_k} \del^2_{jk}V
-\frac{p_j' d}{2h (d - 1)}\ H_h'
+\frac{1}{h^{5/2} (d - 1) x^2} \biggl(
 h' p_j' x^2 h^{1/2}d \\
 &- h^2 \del_jV (d - 1)\biggr)\ H_h\,,
\end{split}
\eqlabel{fin1}
\end{equation}
along with the algebraic expressions for $H_h'$ and $H_h$: 
\begin{equation}
\begin{split}
&H_h'=\frac{2  h(d-1)}{d}\ \sum_{j=1}^p \hp  p_j' 
+ \frac{x (d + 3) h' + 4 (d - 1) h}{4 x  h}\ H_h\,,
\end{split}
\eqlabel{fin2}
\end{equation}
\begin{equation}
\begin{split}
&H_h=-16 h^2 (d - 1)\ 
\biggl[
-120 \left(d - \frac43\right) x^2 h^2\ \sum_{j=1}^p (p_j')^2
+h' xd(d + 1) (17d - 22) (h'x + 8h)
\\&
+(272 d - 352) V h^{5/2}
-16 x^2 d(17 d^2 + 2 d\gamma - 2\gamma^2 - 3d - 20) h^3 \\&+ 16d(d + 1)(17d - 22) h^2
\biggr]^{-1}\
\biggl(
4 h^{3/2}\ \sum_{j=1}^p\hp\del_jV  -4 x^2 h\ \sum_{j=1}^p p_j' \hp'\\
&+x (h' x + 4 h) (d+1)\ \sum_{j=1}^p
\hp p_j' 
\biggr)\,.
\end{split}
\eqlabel{fin3}
\end{equation}

}
}%
}
\bigskip

Equations \eqref{fin1}-\eqref{fin3} is our main result. 

Some comments are in order:
\begin{itemize}
\item It is convenient to use the radial coordinate 
\begin{equation}
x\in (0,\infty)\,,
\eqlabel{xfgr}
\end{equation}
where $x\to 0$ is the location of the asymptotic AdS$ _{d+2}$ boundary, and
$x\to \infty$ corresponding to (see \eqref{fg1}) $c_1^2\to 0$. Special care must be taken to
insure that the bulk geometry is smooth in the latter limit:
\begin{equation}
h=x^{-2}\ \biggl(\frac 14+\calo(x^{-1})\biggr)\,,
\eqlabel{hir}
\end{equation}
see Appendix B.1 of \cite{Buchel:2019pjb} for additional details.
\item To construct the background, it is sufficient to use equations \eqref{fgbac1} and \eqref{fgbac2};
the remaining equation \eqref{fgbac3} is redundant.
\item In applications, we eliminate $H_h'$ and $H_h$ from \eqref{fin1}, using \eqref{fin2} and
\eqref{fin3}.
The master fluctuation equations \eqref{fin1} are solved subject to the following boundary conditions:
\nxt the radial functions $\hp$ are normalizable as $x\to 0$;
\nxt the radial functions $\hp$ have a singularity as $x\to \infty$ as
\begin{equation}
\hp\ \propto (1+2x)^{\gamma/2}\ \times\ {\rm finite}\,.
\eqlabel{singf}
\end{equation}
To understand \eqref{singf}, recall that the bulk boundary conditions
for a black brane QNMs are regular in the EF coordinate system, and are
singular (the incoming-wave) in the FG coordinate system. The difference is
simply due to the difference between the harmonic time-dependence of the linearized
fluctuations in the EF and the FG coordinate systems. Exactly the same story is
applicable here. The difference  between the EF time $t$ and the FG time $\tau$
is \eqref{ttau}
\begin{equation}
t-\tau=\int_{r}^{\infty} \frac{dz}{2A(z)}=-\int_0^\rho h^{1/2}(\rho) d\rho=-\frac 1H\
\int_0^x h^{1/2}(z) dz\,,
\eqlabel{ttaufg}
\end{equation}
where we used the fact, compare \eqref{effgframe} and \eqref{ef2fg}, that
\begin{equation}
\underbrace{\frac{1}{2A(r)}}_{c_1(\rho)^{-2}=\rho^{2}h^{1/2}}\
dr^2 = c_3(\rho)^2\ d\rho^2= \frac{h^{1/2}}{\rho^2}\ d\rho^2\qquad \Longleftrightarrow\qquad
dr=-\frac{d\rho}{\rho^2}\,.
\eqlabel{rrho}
\end{equation}
Thus,
\begin{equation}
e^{-i \omega t} = e^{-i \omega\tau}\ \times\ \exp\biggl\{i\ \frac{\omega}{H}\ \int_0^xh^{1/2}(z) dz \biggr\}
= e^{-i \omega\tau}\ \times\ \exp\biggl\{\gamma\ \int_0^xh^{1/2}(z) dz \biggr\}\,.
\eqlabel{singfg}
\end{equation}
Given \eqref{hir}, as $x\to \infty$,
\begin{equation}
\exp\biggl\{\gamma\ \int_0^xh^{1/2}(z) dz \biggr\}\ \sim\ \exp\biggl\{ \frac\gamma2 \ln x
+ {\rm finite}\biggr\}\ \sim\ x^{\gamma/2}\ \times {\rm finite}\,,
\end{equation}
which explains \eqref{singf} --- we modified $x\to (1+2x)$ to avoid introduction of the
spurious singularity near the AdS$ _{d+2}$ boundary, \ie as $x\to 0$.
\end{itemize}

\section{Applications}\label{examples}

In this section we present various examples to help the reader understand the
computational framework.

\subsection{Probe-limit fluctuations of $\chi\ \sim\ \calo_\Delta$ in $AdS_5$}

Consider effective action \eqref{ea} in $d=3$, with a single free bulk scalar
$\phi_1\equiv \chi$
of mass
\begin{equation}
m_\chi^2 L^2= \Delta (\Delta-4)\,.
\eqlabel{m2}
\end{equation}
This scalar is dual to an operator of dimension $2\le \Delta$ of the boundary theory. 
We consider the model \eqref{ea} in the probe approximation, \ie we neglect the
backreaction of $\chi$ on AdS$ _5$-de Sitter-sliced bulk geometry.
As explained in \cite{Buchel:2021ihu,Buchel:2017pto} this is
a degenerate case of the DFP, since a conformal de Sitter dynamics is Weyl equivalent to
Minkowski dynamics. Nonetheless, this example is instructive and extracts
the leading-order results in the near-conformal limit.  

Solving \eqref{fgbac2} with $p_1(x)=0$, $V=-12$ we find
\begin{equation}
h=\frac{1}{(1+2 x)^2}\,.
\eqlabel{hads}
\end{equation}
The master equation for the $\chi$-fluctuations \eqref{fin1} takes the form
(we relabel $H_{p_1}$ in \eqref{defflfg} as $H_\chi$) :
\begin{equation}
0=H_\chi'' - \frac{x + 3}{(1 + 2 x) x}\ H_\chi'
-\frac{\gamma ( \gamma-3) x^2 + \Delta (\Delta - 4) (1 + 2 x)}{(1 + 2 x)^2 x^2}\ H_\chi\,.
\eqlabel{masprobe}
\end{equation}
Eq.~\ref{masprobe} is solved subject to the following boundary conditions:
\begin{equation}
\begin{split}
&x\to 0:\qquad H_\chi= x^\Delta (1+\calo(x))\,,\\
&x\to \infty:\qquad H_\chi= (1+2x)^{\gamma/2}\ \times \calo(1)\,.
\end{split}
\eqlabel{bcprobe}
\end{equation}
The general solution of \eqref{masprobe} takes form
\begin{equation}
\begin{split}
H_\chi=(1+2x)^{-\gamma/2}\
\biggl\{\ &\cala_1\ x^\Delta\  _2 F_1\left(\Delta-\gamma,\Delta-\frac 32; 2 \Delta-3; -2x\right)
\\
&+\cala_2\ x^{4-\Delta}\   _2 F_1\left(4-\Delta-\gamma,\frac 52-\Delta; 5-2 \Delta; -2x\right)
\biggr\}\,.
\end{split}
\eqlabel{gensolprobe}
\end{equation}
The boundary normalizability of the fluctuations, \ie the first boundary condition in
\eqref{bcprobe}, requires $\cala_1=1$ and $\cala_2=0$. The second boundary condition
in \eqref{bcprobe} determines the spectrum:
\begin{equation}
\gamma=\Delta + k\,,\qquad k\in \zet_+\qquad \Longrightarrow\qquad
\omega =-i H \times \{\Delta,\Delta+1,\cdots\}\,.
\eqlabel{probespectrum}
\end{equation}
Precisely the same spectrum can be obtained in the EF framework of section \ref{efmaster}. 

\subsection{Probe-limit fluctuations of $\chi\ \sim\ \calo_\Delta$ in mass-deformed $AdS_5$}
\label{3.2}

We now consider  a $d=3$  model \eqref{ea} with two scalar fields $\phi$ and $\chi$  
and the potential
\begin{equation}
V=-12 -\frac 32 \phi^2 +\frac{\Delta(\Delta-4)}{2}\ \chi^2\,.
\eqlabel{pot2}
\end{equation}
We turn on the source term for the bulk scalar $\phi$, dual to the  operator of the
 conformal dimension $\Delta_\phi=3$,
\begin{equation}
\phi= \underbrace{\frac{m}{H}}_{p_1}\ x+\calo(x^2)\,,\qquad {\rm as}\qquad x\to 0\,,
\eqlabel{source}
\end{equation}
while we continue to treat the bulk scalar $\chi$ (dual to the operator $\calo_\chi$ of the
conformal dimension $\Delta$) in the probe approximation. We would like to determine the
 spectrum of linearized fluctuations of $\chi$. From the boundary QFT$ _4$  perspective,
we are interested in the $\zet_2$ spontaneous symmetry breaking fluctuations
$\calo_\chi\leftrightarrow - \calo_\chi$ of the massive holographic theory in dS$ _4$.

We outline computational details of the spectrum in the FG coordinate system only ---
the EF computations proceed following the framework of section \ref{efmaster}. 
\begin{itemize}
\item First, we numerically compute the background geometry, solving
the background equations \eqref{fgbac1} and \eqref{fgbac2} for $\phi=p(x)$ and $h(x)$
with the following boundary conditions:
\nxt $x\to 0$, \ie near the AdS$ _5$ asymptotic boundary,
\begin{equation}
\begin{split}
&p=p_1 x+\frac 14 p_1 h_1\ x^2+\left(p_3+\frac16 p_1 (p_1^2+6) \ln x\right)\ x^3+\calo(x^4\ln x)\,,\\
&h=1 + h_1 x + \left(2 + \frac13 p_1^2 + \frac58 h_1^2\right) x^2 + \frac{h_1}{16}
(5 h_1^2 + 8 p_1^2 + 48)\ x^3+\calo(x^4\ln x)\,;
\end{split}
\eqlabel{uv1}
\end{equation}
\nxt $y\equiv \frac 1x\to 0$,
\begin{equation}
p=p^h_0-\frac 35 p^h_0\ y+\calo(y^2)\,,\qquad y^{-2} h=\frac14 - \left(
\frac 14 + \frac{1}{32} (p^h_0)^2\right)\ y+\calo(y^2)\,.
\eqlabel{ir1}
\end{equation}
\\
Note, that for a fixed source $p_1$, the asymptotic expansions are specified with
$\{p_3,h_1,p^h_0\}$ --- the three parameters needed to determine a solution
of a coupled system of a pair of the  second-order \eqref{fgbac1} and the first order
\eqref{fgbac2} ODEs.
\item Next, we solve the master equation \eqref{fin1} for $\chi$. We find it convenient to
represent
\begin{equation}
H_\chi=(1+2x)^{\gamma/2}\ \biggl(\frac{x}{1+x}\biggr)^\Delta\ \hat{H}_\chi\,,
\eqlabel{redefchi}
\end{equation}
so that the asymptotic expansions take the form:
\nxt  $x\to 0$, \ie near the AdS$ _5$ asymptotic boundary,
\begin{equation}
\begin{split}
&\hat{H}_\chi=1 + \left(\frac {\Delta}{4} h_1 + \Delta - \gamma\right)\ x+\calo(x^2)\,,
\end{split}
\eqlabel{uv2}
\end{equation}
\nxt  $y\equiv \frac 1x\to 0$,
\begin{equation}
\begin{split}
&\hat{H}_\chi=c^h_0\biggl(1
+ \frac{(p^h_0)^2\gamma^2 + \left(-\frac{11}{2}(p^h_0)^2
+ 32\Delta - 24\right)\gamma - 16\Delta(\Delta + 1)}
{32\left(\gamma - \frac52\right))}\ y+\calo(y^2)\biggr)\,.
\end{split}
\eqlabel{ir2}
\end{equation}
Note that, as required, a single second-order master equation for $\hat{H}_\chi$
is determined by two parameters $\{\gamma,c^h_0\}$. 
\end{itemize}

\begin{figure}[t]
\begin{center}
\psfrag{p}{{$\frac{m}{H}$}}
\psfrag{d}[t]{{${\gamma_{FG}}/{\gamma_{EF}}-1$}}
\includegraphics[width=4in]{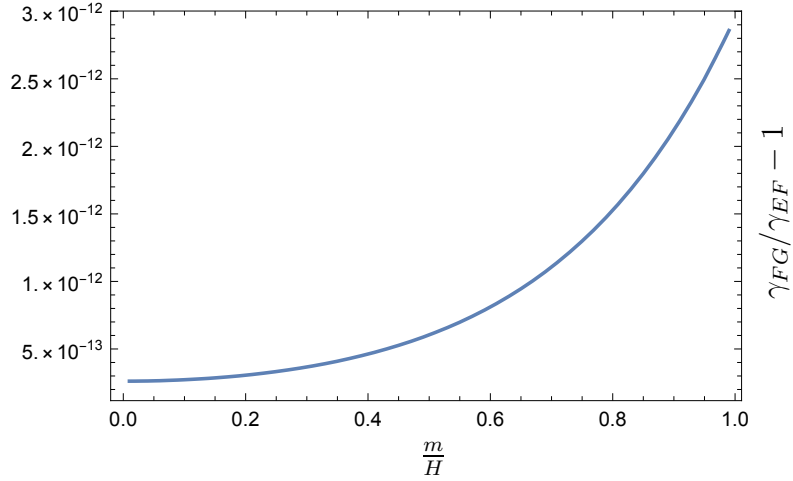}
\end{center}
  \caption{
The spectrum of $\zet_2$ symmetry breaking fluctuations
in model \eqref{pot2}
computed in the FG coordinate system, using the master equations formalism, $\gamma_{FG}$,
is compared to the EF coordinate system computations, $\gamma_{EF}$.
We considered $\Delta=3$ and displayed the mode with $\lim_{p_1\to 0} \gamma =\Delta$. 
} \label{figure1}
\end{figure}

In fig.~\ref{figure1} we present the difference for the spectrum
of $\zet_2$ symmetry breaking fluctuations
in model \eqref{pot2} (we set $\Delta=3$ and consider
the mode with $\lim_{p_1\to 0} \gamma =\Delta$)
computed in the FG coordinate system, using the master equations formalism, $\gamma_{FG}$,
and in the EF coordinate system, $\gamma_{EF}$.

\subsection{Fluctuations in mass-deformed $AdS_5$ beyond the probe
approximation}\label{3.3}

In this section we consider  a $d=3$  model \eqref{ea} with a single scalar fields $\phi$   
and the potential
\begin{equation}
V=-12 -\frac 32 \phi^2\,.
\eqlabel{pot3}
\end{equation}
We turn on the source term for the bulk scalar $\phi$, dual to the  operator of the
 conformal dimension $\Delta=3$,
\begin{equation}
\phi= \underbrace{\frac{m}{H}}_{p_1}\ x+\calo(x^2)\,,\qquad {\rm as}\qquad x\to 0\,.
\eqlabel{source2}
\end{equation}

Once gain, we outline computational details of the spectrum in the FG coordinate system only.
\begin{itemize}
\item The computation of the background geometry is exactly as in section \ref{3.2}. 
\item Next, we solve the master equation \eqref{fin1}. We find it convenient to
represent
\begin{equation}
H_p=(1+2x)^{\gamma/2}\  \hat{H}_p\,,
\eqlabel{redefp}
\end{equation}
so that the asymptotic expansions take the form
\nxt  $x\to 0$, \ie near the AdS$ _5$ asymptotic boundary,
\begin{equation}
\begin{split}
&\hat{H}_p=x^3 +\biggl(-\gamma + \frac34 h_1\biggr)\ x^4 +\calo(x^5\ln x)\,;
\end{split}
\eqlabel{uv3}
\end{equation}
\nxt  $y\equiv \frac 1x\to 0$,
\begin{equation}
\begin{split}
&\hat{H}_\chi=H^h_{p,0}\biggl(
1 + \frac{\gamma (p^h_0)^2(2\gamma - 11) - 48\gamma + 96}{32(2\gamma - 5)}\ y+
\calo(y^2)\biggr)\,.
\end{split}
\eqlabel{ir4}
\end{equation}
Note that, as required, a single second-order master equation for $\hat{H}_p$
is determined by two parameters $\{\gamma,H^h_{p,0}\}$. 
\end{itemize}

\begin{figure}[t]
\begin{center}
\psfrag{p}{{$\frac{m}{H}$}}
\psfrag{d}[t]{{$\gamma=-\Im \frac{\omega}{H}$}}
\includegraphics[width=4in]{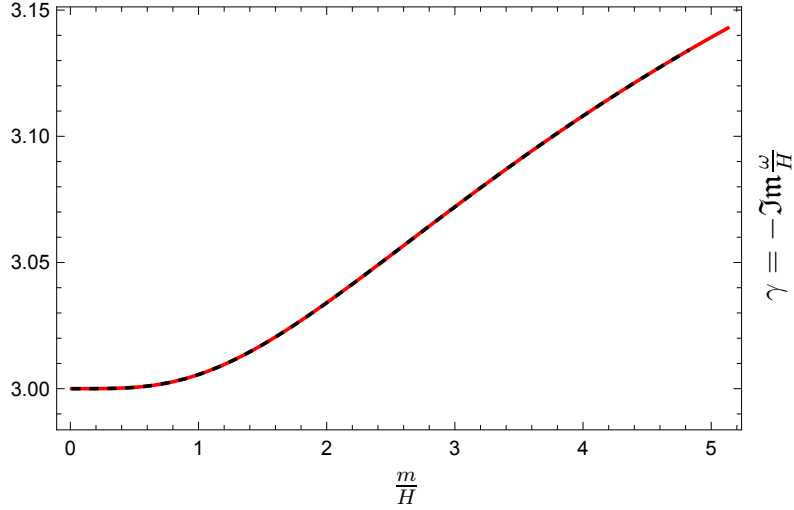}
\end{center}
  \caption{
The lowest lying mode in the spectrum of linearized fluctuations
about the de Sitter DFP of model \eqref{pot3} from
master equation \eqref{fin1} (the black dashed curve) and using the
EF coordinate system of section \ref{efmaster} (the solid red curve).
} \label{figure2}
\end{figure}

In fig.~\ref{figure2} we compare the frequency of the lowest lying mode
of the de Sitter DFP of model \eqref{pot3} computed from the master equation
\eqref{fin1} (the dashed black curve), and using the EF coordinate system formalism of section
\ref{efmaster} (the solid red curve). There is an excellent agreement.

\subsection{Spectrum computed in \cite{Buchel:2017lhu} from the master equation}\label{old}

We use the master equation formalism to reproduce the spectrum of fluctuations of the model
introduced in \cite{Buchel:2017lhu}. Here, we have $d=2$ and a single scalar field $\phi$
with a potential
\begin{equation}
V=-6-\phi^2\,.
\eqlabel{3dpot}
\end{equation}

We begin solving the model in the near conformal limit, \ie
perturbatively in
\begin{equation}
p_1\equiv \frac{\Lambda}{H}
\eqlabel{def1}
\end{equation}
where $p_1$ is the source for the bulk scalar:
\begin{equation}
\phi=p_1\ x+\calo(x^2)\,,\qquad {\rm as}\qquad  x\to 0\,.
\eqlabel{p1def}
\end{equation}
Substituting the perturbative expansions 
\begin{equation}
\phi=\sum_{n=0}^\infty p_1^{2n+1}\ \phi_{2n+1}(x)\,,\qquad h=\frac{1}{(1+2x)^2}
\biggl(\ 1+\sum_{n=1}^\infty p_1^{2n}\ h_{2n}(x)\ \biggr)\,,
\eqlabel{pertph}
\end{equation}
into the background equations \eqref{fgbac1} and \eqref{fgbac2},
and solving them order-by-order in $p_1$ we find:
\begin{equation}
\begin{split}
&\phi_1=\frac{x}{1+2 x}\,,\qquad h_2=-\frac{x}{6(1 + 2 x)^2}\,,\qquad 
\phi_3=-\frac{x^2 (x + 1)}{9(1 + 2 x)^3}\,,\\
&h_4=\frac{x (32 x^2 + 50 x + 5)}
{864(1 + 2 x)^4}\,,\qquad
\phi_5=\frac{x^2 (392 x^3 + 784 x^2 + 468 x + 51)}{12960(1 + 2 x)^5}\,,\\
&h_6=-\frac{x (3776 x^4 + 10432 x^3 + 8784 x^2 + 1794 x + 129)}
{311040(1 + 2 x)^6}\,.
\end{split}
\eqlabel{perres}
\end{equation}
Next, we solve the master equation \eqref{fin1}, using the perturbative expansions
\begin{equation}
H_p=(1+2 x)^{\gamma/2}\ \sum_{n=0}^\infty p_1^{2n} H_{p,2n}(x)\,,\qquad \gamma=\sum_{n=0}^\infty
p_1^{2n}\ \gamma_{2n}\,.
\eqlabel{php1}
\end{equation}
For the lowest lying mode, \ie $\gamma_0=\Delta=2$ we find:
\begin{equation}
\begin{split}
&H_{p,0}=\frac{4 x^2}{(1 + 2 x)^2}\,,\ \
H_{p,2}=-\frac{x^2 (4 x+1)}{3(1+2 x)^4}\,,\ \ 
H_{p,4}=-\frac{x^2 (380 x^3+472 x^2+381 x+84)}{1296(1+2 x)^6}\,,\\
&H_{p,6}=
\frac{x (2216016 x^6+5601888 x^5+5680020 x^4+2966080 x^3+741372 x^2
+69101 x}{4665600(1+2 x)^8}\,,
\end{split}
\eqlabel{sethp2n}
\end{equation}
and
\begin{equation}
\gamma=2 + \frac{1}{12}\ p_1^2 - \frac{1}{54}\ p_1^4 + \frac{1591}{622080}\ p_1^6+\calo(p_1^8)\,,
\eqlabel{w2d2}
\end{equation}
in perfect agreement with the results reported in \cite{Buchel:2017lhu}.

It is straightforward to extend the computations to finite $p_1$; we find the perfect agreement.

We would like to comment on the momentum dependence  of linearized fluctuations
about the DFP. As stated in the introduction, we expect the momentum dependence to be 
trivial in the late time limit. Using the results of \cite{Penin:2021sry},
the non-zero momentum linearized fluctuation $\Phi$ 
has the time dependence as
\begin{equation}
\Phi\ \propto \eta\ J_\nu (k\eta)\,,
\eqlabel{skend1}
\end{equation}
where $\eta$ is the conformal time parameter
\begin{equation}
\eta\equiv -\frac 1H\ e^{-H t}\,,
\eqlabel{conftime}
\end{equation}
and the Bessel index $\nu=\gamma-1$. The late time $t\ \to \infty$ limit corresponds to
$\eta\to 0_-$. It is natural to define a time and momentum dependent
frequency as
\begin{equation}
\underbrace{-i\omega(t)}_{-H \Gamma(t)}\ \equiv \frac{d}{dt}\ \ln\Phi = -H\ \eta\frac{d}{d\eta}\
\ln\Phi\,,
\eqlabel{wdef}
\end{equation}
resulting in
\begin{equation}
\Gamma(t)\equiv \underbrace{(1 + \nu)}_{\gamma} - \frac{(k\eta)^2}{2(1 + \nu)} - \frac{(k\eta)^4}{8(1 + \nu)^2(2 + \nu)}
+\calo\left((k\eta)^6\right)\,,
\eqlabel{gt}
\end{equation}
\ie the finite-momentum decay-rate $\Gamma$
approaches the zero momentum decay-rate 
at late times exponentially fast, 
\begin{equation}
\Gamma(t)-(1+\gamma)\ \propto\ \frac{k^2}{H^2}\ e^{-2 H t}\,,\qquad {\rm as}\qquad t\to \infty\,.
\eqlabel{skdiff}
\end{equation}

\section*{Acknowledgments}
I am indebted to Kostas Skenderis for explanation of \cite{Penin:2021sry}.
I would like to thank the Centro de Ciencias de Benasque Pedro Pascual for hospitality
where part of this work was done.
Research at Perimeter Institute is supported in part by the Government of Canada through the
Department of Innovation, Science and Economic Development Canada and by the
Province of Ontario through the Ministry of Colleges and Universities. This work is further
supported by a Discovery Grant from the Natural Sciences and Engineering
Research Council of Canada.

\appendix
\section{EF frame equations of motion}
\label{efframe}

Within Eddington-Finkelstein metric ansatz (with spatially homogeneous and isotropic background
metric of the $\qft$ --- $ d\boldsymbol{x}^2$)
\begin{equation}
ds_{d+2}^2 = 2dt\ \left(dr-A\ dt\right)+\Sigma^2\ d\boldsymbol{x}^2\,,
\eqlabel{ef1}
\end{equation}
with
\begin{equation}
\begin{split}
A=A(t,r)\,,\qquad \Sigma=\Sigma(t,r)\,,\qquad \phi_j=\phi_j(t,r)\,,
\end{split}
\eqlabel{ef2}
\end{equation}
we find from \eqref{ea} the following evolution equations (\ $'\equiv \del_r$,
$d_+\equiv \del_t+ A\del_r$ and $V_j\equiv\frac{\del V}{\del\phi_j}$ \ ): 
\begin{equation}
\begin{split}
&0=\left(d_+\phi_j\right)'+\frac{d}{2}\ (\ln\Sigma)'\ d_+\phi_j+\frac{d}{2}\ (\phi_j)'\
d_+\ln\Sigma -\frac {V_j}{2} \,,
\end{split}
\eqlabel{ev1}
\end{equation}
\begin{equation}
\begin{split}
&0=\left(d_+\Sigma\right)'+(d-1) (\ln\Sigma)'\ d_+\Sigma+\frac{V}{2d}\ \Sigma\,,
\end{split}
\eqlabel{ev2}
\end{equation}
\begin{equation}
\begin{split}
&0=A''-d(d-1)(\ln\Sigma)'\ d_+\ln\Sigma+\frac12 \sum_{j=1}^p(\phi_j)'d_+\phi_j-\frac{d-2}{2d}V\,,
\end{split}
\eqlabel{ev3}
\end{equation}
and the constraint equations,
\begin{equation}
\begin{split}
&0=\Sigma''+\frac{\Sigma}{2d}\ \sum_{j=1}^p(\phi_j')^2\,,
\end{split}
\eqlabel{con1}
\end{equation}
\begin{equation}
\begin{split}
&0=d_+^2\Sigma-A'\ d_+\Sigma+\frac{\Sigma}{2d}\ \sum_{j=1}^p \left(d_+\phi_j\right)^2\,.
\end{split}
\eqlabel{con2}
\end{equation}

\section{FG frame equations of motion}
\label{fgframe}

For the Fefferman-Graham metric
ansatz (with spatially homogeneous and isotropic background
metric of the $\qft$ --- $ d\boldsymbol{x}^2$) we take 
\begin{equation}
ds_{d+2}^2 =-c_1^2\ d\tau^2 +c_2^2\ d\boldsymbol{x}^2 +c_3^2\ d\rho^2\,,
\eqlabel{fg1}
\end{equation}
with
\begin{equation}
\begin{split}
&c_1\equiv \frac{\sqrt{G_{tt}}}{\rho h^{1/4}}\,,\qquad c_2\equiv
\frac{\sqrt{G_{xx}}}{\rho h^{1/4}}\,,\qquad c_3\equiv \frac{h^{1/4}}{\rho}\,, \\
&G_{tt}=G_{tt}(\tau,\rho)\,,\qquad G_{xx}=G_{xx}(\tau,\rho)\,,\qquad
h=h(\tau,\rho)\,,\qquad
\phi_j=\phi_j(\tau,\rho)\,.
\end{split}
\eqlabel{fg2}
\end{equation}
We find from \eqref{ea} the following equations of motion (\ $\dot{ }\equiv \del_\tau$,
$'\equiv \del_\rho$ and $V_j\equiv\frac{\del V}{\del\phi_j}$\ ): 
\begin{equation}
\begin{split}
&0={h}''-\frac{2 h}{G_{xx}}\ G_{xx}''-\frac{h^2}{G_{tt} d}\
\sum_{j=1}^p \left(\dot{\phi_j}\right)^2-\frac hd\
\sum_{j=1}^p \left({\phi_j'}\right)^2-\frac{h (d-3)}{2G_{xx}^2}\ (G_{xx}')^2
\\&+\frac{h^2 (d-1)}{2G_{xx}^2 G_{tt}}\ (\dot{G}_{xx})^2
-\frac{d+11}{8h} (h')^2+\frac{(d-3)}{8G_{tt}} (\dot{h})^2
+\biggl(
\frac{d+2}{2G_{xx}} G_{xx}'-\frac{d+1}{\rho}\biggr)
h'
\\&-\frac{h(d-2)}{2G_{xx} G_{tt}} \dot{G}_{xx} \dot{h}
+\frac{2 d h}{\rho G_{xx}} G_{xx}'-\frac{2 h^{3/2}}{d \rho^2} V-\frac{2 h (d+1)}{\rho^2}\,,
\end{split}
\eqlabel{eq1}
\end{equation}
\begin{equation}
\begin{split}
&0=h''-\frac{(d-2) h}{G_{tt} d}\ \ddot{h}-\frac{2 h (d-1)}{G_{xx} d}\ G_{xx}''
+\frac{2 h^2 (d-1)}{G_{tt} G_{xx} d}\ \ddot{G}_{xx}-\frac{2 h}{G_{tt} d}\ G_{tt}''
+\frac{d^2+3 d-8}{8G_{tt} d} (\dot{h})^2
\\&-\frac{d+11}{8h} (h')^2+\frac{h}{G_{tt}^2 d} (G_{tt}')^2
+\frac{h^2 (d-1) (d-4)}{2G_{tt} G_{xx}^2 d} (\dot{G}_{xx})^2
-\frac{h (d-1) (d-4)}{2G_{xx}^2 d} (G_{xx}')^2\\&-\frac hd\ \sum_{j=1}^p (\phi_j')^2
+\frac{h^2}{G_{tt} d}\ \sum_{j=1}^p (\dot{\phi_j})^2 
+\biggl(
\frac{d+2}{2dG_{tt}} h'-\frac{h(d-1)}{G_{tt} G_{xx}d} G_{xx}'
+\frac{2 h}{G_{tt} \rho}
\biggr) G_{tt}'
\\&+\biggl(
\frac{h(d-2)}{2G_{tt}^2 d}  \dot{h}-\frac{h^2(d-1)}{G_{xx} G_{tt}^2d} \dot{G}_{xx}
\biggr) \dot{G}_{tt}
-\frac{h(d^2-3 d+2)}{2G_{xx} G_{tt} d} \dot{h}  \dot{G}_{xx}
-\frac{2 h^{3/2}}{d r^2} V
\\&+\biggl(
 \frac{(d+2)(d-1)}{2G_{xx} d} h'
-\frac{2 h(d-1)}{\rho G_{xx}}\biggr) G_{xx}'
-\frac{2 (d+1) h}{\rho^2}- \frac{d+1}{\rho} h'\,,
\end{split}
\eqlabel{eq2}
\end{equation}
\begin{equation}
\begin{split}
&0=\phi_j''-\frac{h}{G_{tt}}\ \ddot{\phi_j}
+\biggl(\frac{d-2}{4G_{tt}} \dot{h}
-\frac{hd}{2G_{xx} G_{tt}} \dot{G}_{xx} 
+\frac{h}{2G_{tt}^2} \dot{G}_{tt}\biggr) \dot{\phi_j}
+\biggl(
\frac{d}{2G_{xx}} G_{xx}'+\frac{1}{2G_{tt}} G_{tt}'\\
&-\frac d\rho
-\frac{d+2}{4h} h'\biggr) \phi_j'
-\frac{h^{1/2}}{\rho^2} \del_j V\,,
\end{split}
\eqlabel{eq3}
\end{equation}
along with the constraints
\begin{equation}
\begin{split}
&0=\ddot{G}_{xx}- \frac{G_{xx}}{2h}\  \ddot{h}
+\frac{G_{xx}}{2d}\  \sum_{j=1}^p (\dot{\phi_j})^2+\frac{G_{xx} G_{tt}}{2h d}\
\sum_{j=1}^p ({\phi_j'})^2
-\frac{G_{tt} (d-1)}{4h G_{xx}} (G_{xx}')^2
\\&+\frac{d-3}{4G_{xx}} (\dot{G}_{xx})^2
+\frac{G_{xx} (d+7)}{16h^2} (\dot{h})^2
-\frac{G_{xx} G_{tt} (d+1)}{16h^3} (h')^2
+\frac{G_{xx}}{4h G_{tt}} \dot{h} \dot{G}_{tt}
-\frac{G_{xx} G_{tt}}{h^{1/2}  \rho^2d} V\\&+\frac{G_{xx} (h' \rho+4 h)}{4h^2 \rho} G_{tt}'
+\biggl(
\frac{G_{tt} d}{4h^2} h'-\frac{1}{2h} G_{tt}'+\frac{G_{tt} d}{h \rho}
\biggr) G_{xx}'
-\biggl(
\frac{1}{2G_{tt}} \dot{G}_{tt}+\frac{d}{4h} \dot{h}\biggr) \dot{G}_{xx}
\\&-\frac{G_{xx} G_{tt} (d+1)}{2h^2 \rho} h'-\frac{(d+1) G_{xx} G_{tt}}{h \rho^2}\,,
\end{split}
\eqlabel{eqc1}
\end{equation}
\begin{equation}
\begin{split}
&0=\dot{h}'-\frac{2 h}{G_{xx}}\ \dot{G}_{xx}'-\frac{2 h}{d}\ \sum_{j=1}^p \dot{\phi_j}\phi_j'
+\frac{h}{G_{xx}} \biggl(\frac{G_{tt}'}{G_{tt}}+\frac{G_{xx}'}{G_{xx}}\biggr) \dot{G}_{xx}
-\frac{5}{4h} \dot{h} h'
-\frac{1}{2G_{tt}} \dot{h} G_{tt}'\\&+\frac{1}{G_{xx}} \dot{h} G_{xx}'-\frac 1\rho \dot{h}\,.
\end{split}
\eqlabel{eqc2}
\end{equation}

\bibliographystyle{JHEP}
\bibliography{masterds}

\providecommand{\href}[2]{#2}\begingroup\raggedright\begin{thebibliography}{10}

\bibitem{Buchel:2017lhu}
A.~Buchel, \emph{{Ringing in de Sitter spacetime}},
  \href{http://dx.doi.org/10.1016/j.nuclphysb.2018.01.021}{\emph{Nucl. Phys.}
  {\bf B928} (2018) 307--320}, [\href{https://arxiv.org/abs/1707.01030}{{\tt
  1707.01030}}].

\bibitem{Buchel:2021ihu}
A.~Buchel, \emph{{Dynamical fixed points in holography}},
  \href{http://dx.doi.org/10.1007/JHEP02(2022)128}{\emph{JHEP} {\bf 02} (2022)
  128}, [\href{https://arxiv.org/abs/2111.04122}{{\tt 2111.04122}}].

\bibitem{Buchel:2017pto}
A.~Buchel and A.~Karapetyan, \emph{{de Sitter Vacua of Strongly Interacting
  QFT}}, \href{http://dx.doi.org/10.1007/JHEP03(2017)114}{\emph{JHEP} {\bf 03}
  (2017) 114}, [\href{https://arxiv.org/abs/1702.01320}{{\tt 1702.01320}}].

\bibitem{Buchel:2017qwd}
A.~Buchel, \emph{{Verlinde Gravity and AdS/CFT}},
  \href{https://arxiv.org/abs/1702.08590}{{\tt 1702.08590}}.

\bibitem{Buchel:2019qcq}
A.~Buchel, \emph{{Entanglement entropy of ${\cal N}=2^*$ de Sitter vacuum}},
  \href{https://arxiv.org/abs/1904.09968}{{\tt 1904.09968}}.

\bibitem{Buchel:2019pjb}
A.~Buchel, \emph{{$\chi\rm{SB}$ of cascading gauge theory in de Sitter}},
  \href{http://dx.doi.org/10.1007/JHEP05(2020)035}{\emph{JHEP} {\bf 05} (2020)
  035}, [\href{https://arxiv.org/abs/1912.03566}{{\tt 1912.03566}}].

\bibitem{Casalderrey-Solana:2020vls}
J.~Casalderrey-Solana, C.~Ecker, D.~Mateos and W.~Van Der~Schee,
  \emph{{Strong-coupling dynamics and entanglement in de Sitter space}},
  \href{http://dx.doi.org/10.1007/JHEP03(2021)181}{\emph{JHEP} {\bf 03} (2021)
  181}, [\href{https://arxiv.org/abs/2011.08194}{{\tt 2011.08194}}].

\bibitem{Ecker:2021cvz}
C.~Ecker, W.~van~der Schee, D.~Mateos and J.~Casalderrey-Solana,
  \emph{{Holographic evolution with dynamical boundary gravity}},
  \href{http://dx.doi.org/10.1007/JHEP03(2022)137}{\emph{JHEP} {\bf 03} (2022)
  137}, [\href{https://arxiv.org/abs/2109.10355}{{\tt 2109.10355}}].

\bibitem{Penin:2021sry}
J.~M. Pen\'\i{}n, K.~Skenderis and B.~Withers, \emph{{Massive holographic QFTs
  in de Sitter}},
  \href{http://dx.doi.org/10.21468/SciPostPhys.12.6.182}{\emph{SciPost Phys.}
  {\bf 12} (2022) 182}, [\href{https://arxiv.org/abs/2112.14639}{{\tt
  2112.14639}}].

\bibitem{WMAP:2003ivt}
{\scshape WMAP} collaboration, C.~L. Bennett et~al., \emph{{First year
  Wilkinson Microwave Anisotropy Probe (WMAP) observations: Preliminary maps
  and basic results}}, \href{http://dx.doi.org/10.1086/377253}{\emph{Astrophys.
  J. Suppl.} {\bf 148} (2003) 1--27},
  [\href{https://arxiv.org/abs/astro-ph/0302207}{{\tt astro-ph/0302207}}].

\bibitem{Kodama:2003jz}
H.~Kodama and A.~Ishibashi, \emph{{A Master equation for gravitational
  perturbations of maximally symmetric black holes in higher dimensions}},
  \href{http://dx.doi.org/10.1143/PTP.110.701}{\emph{Prog. Theor. Phys.} {\bf
  110} (2003) 701--722}, [\href{https://arxiv.org/abs/hep-th/0305147}{{\tt
  hep-th/0305147}}].

\bibitem{Jansen:2019wag}
A.~Jansen, A.~Rostworowski and M.~Rutkowski, \emph{{Master equations and
  stability of Einstein-Maxwell-scalar black holes}},
  \href{http://dx.doi.org/10.1007/JHEP12(2019)036}{\emph{JHEP} {\bf 12} (2019)
  036}, [\href{https://arxiv.org/abs/1909.04049}{{\tt 1909.04049}}].

\bibitem{Buchel:2021ttt}
A.~Buchel, \emph{{Stabilization of the extended horizons}},
  \href{http://dx.doi.org/10.1103/PhysRevD.104.046025}{\emph{Phys. Rev. D} {\bf
  104} (2021) 046025}, [\href{https://arxiv.org/abs/2105.13380}{{\tt
  2105.13380}}].

\bibitem{inprep}
A.~Buchel, \emph{{Fluctuation spectra of the cascading gauge theory de Sitter
  DFPs}}, {\emph{{\rm in preparation}} }.

\bibitem{Chesler:2013lia}
P.~M. Chesler and L.~G. Yaffe, \emph{{Numerical solution of gravitational
  dynamics in asymptotically anti-de Sitter spacetimes}},
  \href{http://dx.doi.org/10.1007/JHEP07(2014)086}{\emph{JHEP} {\bf 07} (2014)
  086}, [\href{https://arxiv.org/abs/1309.1439}{{\tt 1309.1439}}].

\bibitem{Buchel:2014gta}
A.~Buchel, R.~C. Myers and A.~van Niekerk, \emph{{Nonlocal probes of
  thermalization in holographic quenches with spectral methods}},
  \href{http://dx.doi.org/10.1007/JHEP02(2015)017}{\emph{JHEP} {\bf 02} (2015)
  017}, [\href{https://arxiv.org/abs/1410.6201}{{\tt 1410.6201}}].

\end{thebibliography}\endgroup

\end{document}